\documentclass[preprint,aps,prd,nofootinbib]{revtex4}
\usepackage{amsmath}
\usepackage{graphicx}
\usepackage{subfigure}

\newcommand{\bea}{\begin{eqnarray}}
\newcommand{\eea}{\end{eqnarray}}
\newcommand{\beq}{\begin{equation}}
\newcommand{\eeq}{\end{equation}}

\def\/{\over}

\begin{document}

\title{ Spontaneous absorption of an accelerated hydrogen
atom near a conducting plane in vacuum}

\author{ Hongwei Yu$^{1,2}$ and Zhiying Zhu${^2}$ }
\affiliation{ $^1$CCAST(World Lab.), P. O. Box 8730, Beijing,
100080, P. R. China\\
$^2$Department of Physics and Institute of  Physics,\\
Hunan Normal University, Changsha, Hunan 410081, China
\footnote{Mailing address} }
\begin{abstract}
We study, in the multipolar coupling scheme, a uniformly
accelerated multilevel hydrogen atom in interaction with the
quantum electromagnetic field near a conducting boundary and
separately calculate the contributions of the vacuum fluctuation
and radiation reaction to the rate of change of the mean atomic
energy. It is found that the perfect balance between
 the contributions of vacuum fluctuations and radiation reaction
that ensures the stability of ground-state atoms is disturbed,
making spontaneous transition of ground-state atoms to excited
states possible in vacuum with a conducting boundary.  The
boundary-induced contribution is effectively a nonthermal
correction, which enhances or weakens the nonthermal effect
already present in the unbounded case, thus possibly making the
effect easier to observe. An interesting feature worth being noted
is that the nonthermal corrections may vanish for atoms on some
particular trajectories.

\end{abstract}

\maketitle

\baselineskip=16pt

\section{Introduction}

Understanding the physical origin of radiative properties of
atoms, such as spontaneous emission and radiative level shifts, is
a very stimulating problem.  So far mechanisms such as vacuum
fluctuations \cite{Welton48,Compagno83} and radiation reaction
\cite{Ackerhalt73}, or a combination of them \cite{Milonni75},
have been proposed as the possible physical interpretations.   The
ambiguity arises because of the freedom in the choice of ordering
of commuting operators of the atom and field in a Heisenberg
picture approach to the problem. As a result, there exists an
indetermination in the separation of effects of vacuum
fluctuations and radiation reaction such that distinct
contributions of vacuum fluctuations and radiation reaction to the
spontaneous emission of atoms do not possess an independent
physical meaning.  Therefore, although quantitative results for
spontaneous emission and radiative level shifts are
well-established, the physical interpretations remained
controversial until Dalibard, Dupont-Roc and Cohen-Tannoudji(DDC)
argued in \cite{Dalibard82} and \cite{Dalibard84} that there
exists a symmetric operator ordering of atom and field variables
where the distinct contributions of vacuum fluctuations and
radiation reaction to the rate of  change of an atomic observable
are separately Hermitian. If one demands such an ordering, an
independent physical meaning can be assigned to each contribution.
Using this prescription one can show that for ground-state atoms,
the contributions of vacuum fluctuations and radiation reaction to
the rate of change of the mean excitation energy cancel exactly
and this cancellation forbids any transitions from the ground
state and thus ensures atom's stability in vacuum. While for any
initial excited state, the rate of change of atomic energy
acquires equal contributions from vacuum fluctuations and from
radiation reaction.

Recently,  Audretsch, M\"ueller and Holzmann
\cite{Audretsch94,1Audretsch95,2Audretsch95} have generalized the
formalism of DDC \cite{Dalibard84} to evaluate vacuum fluctuations
and radiation reaction contributions to the spontaneous excitation
rate and radiative energy shifts of an accelerated two-level atom
interacting with a scalar field in a unbounded Minkowski space. In
particular, their results show that when an atom is accelerated,
then the delicate balance between vacuum fluctuations and
radiation reaction is altered since the contribution of vacuum
fluctuations to the rate of change of the mean excitation energy
is modified while that of the radiation reaction remains the same.
Thus transitions to excited states for ground-state atoms become
possible even in vacuum. This result not only is consistent with
the Unruh effect \cite{Unruh}, which is closely related to the
Hawking radiation of black holes,  but also provides a physically
appealing interpretation of it, since the spontaneous excitation
of accelerated atoms can be considered as the actual physical
process underlying the Unruh effect. Physically, this gives a
transparent illustration for why an accelerated detector clicks (
See Ref.~\cite{TP85} for a discussion in a different context).

Therefore, one sees that the Unruh effect is intrinsically related
to the effects of modified vacuum fluctuations induced by the
acceleration of the atom (or detector) in question.  On the other
hand, however, it is well-known that the presence of boundaries in
a flat spacetime also modifies the vacuum fluctuations of quantum
fields, and it has been demonstrated that this modification (or
changes) in vacuum fluctuations can lead to a lot of novel
effects, such as the Casimir effect \cite{Casimir}, the light-cone
fluctuations when gravity is quantized \cite{YF99,YF00,YW}, and
the Brownian (random) motion of test particles in an
electromagnetic vacuum \cite{YF04,YC,YCW06,TY} (Also see
\cite{JR,Barton,JR1}), just to name a few. Therefore, it remains
interesting to see what happens to the radiation properties of
accelerated atoms found in Ref.~\cite{Audretsch94} when the vacuum
fluctuations are further modified by the presence of boundaries.
Recently the effects of modified vacuum fluctuations and radiation
reaction due to the presence of a conducting plane boundary upon
the spontaneous excitation of both an inertial and a uniformly
accelerated atom interacting with a quantized real massless scalar
field have been discussed \cite{H. Yu}. It is found that the
modifications induced by the presence of a boundary make the
spontaneous radiation rate of an excited inertial atom to
oscillate near the boundary and this oscillatory behavior may
offer a possible opportunity for experimental tests for
geometrical (boundary) effects in flat spacetime. While for
accelerated atoms, the transitions from ground states to excited
states are found to be possible even in vacuum due to changes in
the vacuum fluctuations induced by both the presence of the
boundary and the acceleration of atoms. Meanwhile the contribution
of radiation reaction is now dependent on the acceleration of the
atom, in sharp contrast to the unbounded Minkowski space where it
has been shown that for accelerated atoms on arbitrary stationary
trajectory, the contribution of radiation reaction is generally
not altered from its inertial value \cite{2Audretsch95}.

However, a two-level atom interacting with a scalar field is more
or less a toy model, and a more realistic system would be a
multi-level atom, a hydrogen atom, for instance, in interaction
with a quantized electromagnetic field. Let us note that such a
system was examined in terms of the radiative energy shifts of an
accelerated atom \cite{Passante97} using the method of
Ref.~\cite{1Audretsch95}, where non-thermal corrections to the
energy shifts were found in addition to the usual thermal ones
associated with the temperature $T=a/2\pi$.  Recently, the
spontaneous excitation rate of an accelerated atom in the same
 system has been studied \cite{ZYL06}. It has been found that both
the effects of vacuum fluctuations and radiation reaction on the
atom are changed by the acceleration. This is in sharp contrast to
the scalar field case where the contribution of radiation reaction
is not altered by the acceleration.  A dramatic feature is that
the contribution of electromagnetic vacuum fluctuations to the
spontaneous emission rate contains an extra non-thermal term
proportional to $a^2$, the proper acceleration squared, in
contrast to the scalar field case where the effect of acceleration
is purely thermal. Therefore the equivalence between uniform
acceleration and thermal fields is lost when the scalar field is
replaced by the electromagnetic field as has been argued elsewhere
in other different context \cite{Boyer,Tak}.  However, one may
wonder what happens to the spontaneous emission of accelerated
multi-level atoms in interaction with quantized electromagnetic
fields found in Ref.~\cite{ZYL06},  when the vacuum fluctuations
are further modified by the presence of boundaries.  This is what
we plan to address in the present paper; we will calculate the
effects of modified vacuum fluctuations and radiation reaction due
to the presence of a conducting plane boundary upon the
spontaneous excitation of both an inertial and a uniformly
accelerated multi-level atom interacting with a quantized
electromagnetic field in the multipolar coupling scheme. It should
be pointed out that the multi-level atom in the dipole coupling
with electromagnetic fields only serves as a model for discussion
and it is still a crude representation of a hydrogen atom in
reality.

The paper is organized as follows, we give, in Sec. II,  a review
of the general formalism developed in Ref.~\cite{Audretsch94} and
generalized in Ref.~\cite{Passante97,ZYL06} to the case of a
multi-level atom interacting with a quantized electromagnetic
field in the multipolar coupling scheme, then apply it to the case
of an inertial atom in Sec. III and to the case of an accelerated
atom in Sec. IV. Finally we will conclude with some discussions in
Sec. V.

\section{The general formalism for vacuum fluctuation and radiation reaction}
We consider a multilevel hydrogen atom in interaction with
electromagnetic fields. To study the modifications of the
spontaneous emission rate of atoms caused by the presence of a
conducting plane boundary in vacuum, we assume that the conducting
boundary is located at $z=0$ in space and consider a pointlike
hydrogen atom on a stationary space-time trajectory $x(\tau)$,
where $\tau$ denotes the proper time on the trajectory. The
stationary trajectory guarantees the existence of a series of
stationary atomic states $|n\rangle$, with energies $\omega_n$.
The Hamiltonian that governs the time evolution of the atom with
respect to the proper time $\tau$ can then be written as
\footnote{Lorentz-Heaviside units with $\hbar=c=1$ will be used
here.}
\begin{eqnarray}
H_A(\tau)=\sum_n\omega_n\sigma_{nn}(\tau)\;,
\end{eqnarray}
where $\sigma_{nn}(\tau)=|n \rangle \langle n|$. The free
Hamiltonian of the quantum electromagnetic field that governs its
time evolution with respect to $\tau $ is
\begin{eqnarray}
H_F(\tau)=\sum_k \omega_{\vec{k}} a_{\vec{k}}^\dag a_{\vec{k
}}{dt\/d \tau} \;,
\end{eqnarray}
where $\vec{k}$ denotes the wave vector and polarization of the
field modes. We couple the hydrogen atom and the quantum
electromagnetic field in the multipolar coupling scheme
\cite{CPP95}
\begin{eqnarray}
H_I(\tau)=-e\textbf{ r}(\tau) \cdot
\textbf{E}(x(\tau))=-e\sum_{mn}\textbf{r}_{mn}\cdot
\textbf{E}(x(\tau))\sigma_{mn}(\tau)\;, \label{HI3}
\end{eqnarray}
where e is the electron electric charge, e$\textbf{r}$ the atomic
electric dipole moment,
$x(\tau)\leftrightarrow(t(\tau),\textbf{x}(\tau))$, the space-time
coordinates of the hydrogen atom. In present case the dipole
moment must be kept fixed with respect to the proper frame of
reference of the atom, otherwise the rotation of the dipole moment
will bring in extra time dependence in addition to the intrinsic
time evolution \cite{Tak}.

Let us note that, since both $\textbf{r}(\tau)$ and
$\textbf{E}(x)$ are not world vectors,  the interaction
Hamiltonian $H_I $ is ambiguous when we deal with the situation of
moving atoms. However, a manifestly coordinate invariant
generalization of $H_I $ can be given \cite{Tak}:
\begin{equation}
H_I(\tau)=-e\,r^{\mu}(\tau)\,F_{\mu\nu}(x(\tau))\,u^{\nu}(\tau)\;,
\;\label{HI4}
\end{equation}
where $F_{\mu\nu}$ is the field strength, $r^{\mu}(\tau)$ is a
four vector such that its temporal component in the frame of the
atom (proper reference frame) vanishes and its spatial components
in the same frame are given by $\textbf{r}(\tau)$, and $u^{\nu}$
is the four velocity of the atom. Since $u^{\nu}(\tau)=(1,0,0,0)$
in the frame of the atom, this extended interaction Hamiltonian
reduces to that given by Eq.~(\ref{HI3}) in the reference frame of
the atom. In what follows, we choose to work in this reference
frame.

 We can now obtain the Heisenberg
equations of motion for the dynamical variables of the hydrogen
atom and the electromagnetic field from the Hamiltonian
$H=H_A+H_F+H_I$. The solutions of the equations of motion can be
split into two parts: a free part, which is present even in the
absence of the coupling, and a source part, which is caused by the
interaction of the atom and field. We assume that the initial
state of the field is the vacuum $|0\rangle$, while the atom is in
the state $|b\rangle$. Our aim is to identify and separate the two
physical mechanisms that contribute to the rate of change of
atomic observables $O(\tau)$: the contribution of vacuum
fluctuations and that of radiation reaction. For this purpose, we
choose a symmetric ordering between atom and field variables and
identify the contribution of the vacuum fluctuations and radiation
reaction to the rate of change of $O(\tau)$. Since we are
interested in the spontaneous emission and absorption of the atom,
we will concentrate on the mean atomic excitation energy $\langle
H_A(\tau)\rangle$. The contribution of vacuum fluctuations(vf) and
radiation reaction(rr) to the rate of change of $\langle
H_A(\tau)\rangle$ can be written as ( cf.
Ref.\cite{Dalibard82,Dalibard84,Audretsch94, ZYL06} )
\begin{eqnarray}
{\biggl\langle
{dH_A(\tau)\/d\tau}\biggr\rangle_{vf}}=2ie^2\int_{\tau_0}^\tau
d\tau' C_{ij}^F(x(\tau),x(\tau')){d\/d\tau}(\chi_{ij}^A)_b
(\tau,\tau')\;,\label{hvf}
\end{eqnarray}
\begin{eqnarray}
{\biggl\langle
{dH_A(\tau)\/d\tau}\biggr\rangle_{rr}}=2ie^2\int_{\tau_0}^\tau
d\tau' \chi_{ij}^F(x(\tau),x(\tau')){d\/d\tau}(C_{ij}^A)_b
(\tau,\tau')\;,\label{hrr}
\end{eqnarray}
where $|\rangle=|b,0\rangle$. The statistical functions of the
atom, $(C_{ij}^A)_b(\tau,\tau')$ and
$(\chi_{ij}^A)_b(\tau,\tau')$, are defined as
\begin{eqnarray}
(C_{ij}^A)_b(\tau,\tau')={1\/2}\langle
b|\{r_i^f(\tau),r_j^f(\tau')\}|b\rangle\;,
\end{eqnarray}\begin{eqnarray}
(\chi_{ij}^A)_b(\tau,\tau')={1\/2}\langle
b|~[r_i^f(\tau),r_j^f(\tau')]|b\rangle\;,
\end{eqnarray}
and those of the field are
\begin{eqnarray}
C_{ij}^F(x(\tau),x(\tau'))={1\/2}\langle0|\{E_i^f(x(\tau)),E_j^f(x(\tau'))\}|0\rangle\label{cf}\;,
\end{eqnarray}
\begin{eqnarray}
\chi_{ij}^F(x(\tau),x(\tau'))={1\/2}\langle0|~[E_i^f(x(\tau)),E_j^f(x(\tau'))]|0\rangle\label{xf}\;.
\end{eqnarray}
Let us note that $C^A$ is also called the symmetric correlation
function of the atom in the state $|b\rangle$, $\chi^A$ its linear
susceptibility, while $C^F$ and $\chi^F$ are also known as the
Hadamard function and Pauli-Jordan or Schwinger function of the
field respectively. The explicit forms of the statistical
functions of the atom are given by
\begin{eqnarray}
(C_{ij}^A)_b(\tau,\tau')={1\/2}\sum_d[\langle
b|r_i(0)|d\rangle\langle d|r_j(0)|b\rangle
e^{i\omega_{bd}(\tau-\tau')}+\langle b|r_j(0)|d\rangle\langle
d|r_i(0)|b\rangle e^{-i\omega_{bd}(\tau-\tau')}]\label{ca}\;,
\end{eqnarray}\begin{eqnarray}
(\chi_{ij}^A)_b(\tau,\tau')={1\/2}\sum_d[\langle
b|r_i(0)|d\rangle\langle d|r_j(0)|b\rangle
e^{i\omega_{bd}(\tau-\tau')}-\langle b|r_j(0)|d\rangle\langle
d|r_i(0)|b\rangle e^{-i\omega_{bd}(\tau-\tau')}]\label{xa}\;,
\end{eqnarray}
where $\omega_{bd}=\omega_b-\omega_d$ and the sum extends over a
complete set of atomic states.

In order to calculate the statistical functions of the field, let
us recall that the two point function for the photon field may be
expressed as
\begin{eqnarray}
D^{\mu\nu}(x,x')=\langle0|A^\mu(x)A^\nu(x')|0\rangle=D_0^{\mu\nu}(x-x')+D_b^{\mu\nu}(x,x')\label{D}\;,
\end{eqnarray}
where $D_0^{\mu\nu}(x-x')$ is the two point function in the usual
Minkowski vacuum, and $D_b^{\mu\nu}(x,x')$, is the correction
induced by the presence of boundary, which can be obtained by the
method of images. In the Feynman gauge, at a distance $z$ from the
boundary, we have, in the laboratory frame,
\begin{eqnarray}
D_0^{\mu\nu}(x-x')={\eta^{\mu\nu}\/{4\pi^2
[(t-t^\prime-i\varepsilon)^2-(x-x^\prime)^2-(y-y^\prime)^2-(z-z^\prime)^2]}}
\end{eqnarray}
and
\begin{eqnarray}
D_b^{\mu\nu}(x,x')=-{{\eta^{\mu\nu}+2n^\mu n^\nu}\/{4\pi^2
[(t-t^\prime-i\varepsilon)^2-(x-x^\prime)^2-(y-y^\prime)^2-(z+z^\prime)^2]}}\;.
\end{eqnarray}
Here, $\eta^{\mu\nu}=diag(1,-1,-1,-1)$ and the unit normal vector
$n^\mu=(0,0,0,1)$. Note that the two-point function Eq.~(\ref{D})
is constructed in such way that the tangential components of the
electric field two-point function vanish on the conducting plane.
The electric field two-point function can be expressed as a sum of
the Minkowski vacuum term and a correction term due to the
boundary:
\begin{eqnarray}
\langle0| \textbf{E}(x)\textbf{E}(x')|0\rangle=\langle0|
\textbf{E}(x)\textbf{E}(x')|0\rangle_0+\langle0|
\textbf{E}(x)\textbf{E}(x')|0\rangle_b\;.
\end{eqnarray}
Since the boundary-independent contributions caused by the
Minkowski vacuum term have been studied in Ref.~\cite{ZYL06}, in
the present paper, we will only calculate the boundary-dependent
contributions, and write
\begin{eqnarray}
\langle0|
E_i(x(\tau))E_j(x(\tau'))|0\rangle_b&=&{1\/4\pi^2}[\,(\delta_{ij}-2n_in_j)\,\partial
_0\partial_0^\prime-\partial_i\partial_j^\prime\,]\nonumber\\&&\times
{1\/(t-t^\prime-i\varepsilon)^2-(x-x^\prime)^2-(y-y^\prime)^2-(z+z^\prime)^2}\;,\label{EER}
\end{eqnarray}
where $\varepsilon\rightarrow+0$ and $\partial^\prime$ denotes the
differentiation with respect to $x^\prime$. The statistical
functions of the field can be calculated using (\ref{EER}).

\section{Spontaneous emission from a uniformly moving atom}
In this section, we apply the previously developed formalism to
study, in the presence of a conducting plane boundary,  the
spontaneous emission of an inertial multilevel atom interacting
with quantized electromagnetic fields in the multipolar coupling
scheme. We consider the atom moving in the $x$-direction with a
constant velocity $v$ at a distance $z$ from the plane, thus its
trajectory is given by
\begin{eqnarray}
t(\tau)=\gamma\tau\;, \ \ \ x(\tau)=x_0+v\gamma\tau\;, \ \ \
y(\tau)=y_0\;,\ \ \ z(\tau)=z\;,
\end{eqnarray}
where $\gamma=(1-v^2)^{-{1\/2}}$. From the general form
Eq.~(\ref{EER}) we can obtain the non-zero electric field
two-point functions in  the frame of the atom \begin{eqnarray}
\langle0|E_x(x(\tau))E_x(x(\tau'))|0\rangle_b&=&\langle0|E_y(x(\tau))E_y(x(\tau'))|0\rangle_b
\nonumber\\&=&-{u^2+4z^2\/\pi^2[\;(u-i\varepsilon)^2-4z^2]^3}\;,\label{EEx}
\end{eqnarray}
and
\begin{eqnarray}
\langle0|E_z(x(\tau))E_z(x(\tau'))|0\rangle_b
={1\/\pi^2[\;(u-i\varepsilon)^2-4z^2]^2}\label{EEz}\;.
\end{eqnarray}
where $u=\tau-\tau'$. Performing calculations using the above
result lead to the non-zero Hadamard functions of the field:
\begin{eqnarray}
C_{xx}^F(x(\tau),x(\tau'))=C_{yy}^F(x(\tau),x(\tau'))=-{1\/2\pi^2}\biggl({u^2+4z^2\/[\;(u-i\varepsilon)^2-4z^2]^3}+
{u^2+4z^2\/[\;(u+i\varepsilon)^2-4z^2]^3}\biggr)\;,\nonumber\\
\end{eqnarray}
\begin{eqnarray}
C_{zz}^F(x(\tau),x(\tau'))={1\/2\pi^2}\biggl({1\/[\;(u-i\varepsilon)^2-4z^2]^2}
+{1\/[\;(u+i\varepsilon)^2-4z^2]^2}\biggr)\;,
\end{eqnarray}
and the Pauli-Jordan, or Schwinger functions:
\begin{eqnarray}
\chi_{xx}^F(x(\tau),x(\tau'))=\chi_{yy}^F(x(\tau),x(\tau'))=-{i\/4\pi
z}\;{u^2+4z^2\/6u^2+8z^2}\biggl(\delta''(u-2z)-\delta''(u+2z)\biggr)\;,
\end{eqnarray}
\begin{eqnarray}
\chi_{zz}^F(x(\tau),x(\tau'))={i\/8\pi z}\;{1
\/u}\biggl(\delta'(u+2z)-\delta'(u-2z)\biggr)\;.
\end{eqnarray}
Here $\delta'$ and $\delta''$ are the first and the second
derivative of the delta function respectively.

With all the statistical functions given, we are ready to
calculate the contributions of both the vacuum fluctuations and
radiation reaction to the rate of change of the mean atomic
energy. Since the polarization direction of the atom can be
arbitrary, in general, the polarization can have non-zero
components in both the direction normal and that which is parallel
to the plane. So calculations have to be carried out for all
non-zero field statistical functions. Take the $xx$ component for
example; it is easy to show that the contribution of the changes
in vacuum fluctuations induced by the presence of the boundary is
given by
\begin{eqnarray}
{\biggl\langle {d H_A(\tau)\/d\tau}\biggr\rangle}_{b,
vf}^{xx}&=&{e^2\/2\pi^2}\sum_d{|\langle
b|r_x(0)|d\rangle|}^2\;\omega_{bd}\nonumber\\&&\times
\int_{-\infty}^\infty
du\biggl({u^2+4z^2\/[\;(u-i\varepsilon)^2-4z^2]^3}+
{u^2+4z^2\/[\;(u+i\varepsilon)^2-4z^2]^3}\biggr)e^{i\omega_{bd}u}\label{vfx1}
\end{eqnarray}
and that of radiation reaction by
\begin{eqnarray}
{\biggl\langle {d H_A(\tau)\/d\tau}\biggr\rangle}_{b,
RR}^{xx}&=&{ie^2\/4\pi z}\sum_d{|\langle
b|r_x(0)|d\rangle|}^2\;\omega_{bd}\nonumber\\&&\times
\int_{-\infty}^\infty
du\;{u^2+4z^2\/6u^2+8z^2}\;\biggl(\delta''(u-2z)-\delta''(u+2z)\biggr)e^{i\omega_{bd}u}\label{rrx1}\;,
\end{eqnarray}
where we have extended the range of integration to infinity for
sufficiently long times $\tau-\tau_0$. The superscript $xx$
denotes contributions associated with the $xx$ component of the
statistical functions and $b$ in the subscript indicates
boundary-dependent contribution. The integrals in Eq.~(\ref{vfx1})
and Eq.~(\ref{rrx1}) can be evaluated via the residue theorem to
get
\begin{eqnarray}
{\biggl\langle {d
H_A(\tau)\/d\tau}\biggr\rangle}_{b,vf}^{xx}&=&{e^2\/32\pi
}\biggl(\sum_{\omega_b>\omega_d}\omega_{bd}^4{|\langle
b|r_x(0)|d\rangle|}^2\,f_x(z,\omega_{bd})
-\sum_{\omega_b<\omega_d} \omega_{bd}^4{|\langle
b|r_x(0)|d\rangle|}^2\,f_x(z,\omega_{bd})\,\bigg)\;,
\nonumber\\\label{vfx2}
\end{eqnarray}
and
\begin{eqnarray}
{\biggl\langle {d
H_A(\tau)\/d\tau}\biggr\rangle}_{b,rr}^{xx}&=&{e^2\/32\pi
}\biggl(\sum_{\omega_b>\omega_d}\omega_{bd}^4{|\langle
b|r_x(0)|d\rangle|}^2\,f_x(z,\omega_{bd})
+\sum_{\omega_b<\omega_d} \omega_{bd}^4{|\langle
b|r_x(0)|d\rangle|}^2\,f_x(z,\omega_{bd})\,\biggr)\,.\nonumber\\
\label{rrx2}
\end{eqnarray}
Here we have defined
\begin{equation}
f_x(z,\omega_{bd})={2\/z^2\omega_{bd}^2}\cos(2z\omega_{bd})+
{{4z^2\omega_{bd}^2-1}\/z^3\omega_{bd}^3}\sin(2z\omega_{bd})
\end{equation}
 Adding up the contributions of vacuum
fluctuations and radiation reaction, we obtain the  rate of change
of the atomic excitation energy induced by the presence of the
boundary.
\begin{eqnarray}
{\biggl\langle {d H_A(\tau)\/d\tau}\biggr\rangle}_{b, tot}^{xx}&=&
{\biggl\langle {d H_A(\tau)\/d\tau}\biggr\rangle}_{b, vf}^{xx}+
{\biggl\langle{dH_A(\tau)\/d\tau}\biggr\rangle}_{b,
rr}^{xx}\nonumber\\&=& {e^2\/16\pi
}\sum_{\omega_b>\omega_d}\omega_{bd}^4{|\langle
b|r_x(0)|d\rangle|}^2\,f_x(z,\omega_{bd})\;.\label{totxR}
\end{eqnarray}
Eq.~(\ref{totxR}) only gives the correction to the spontaneous
excitation rate caused by the presence of boundary and it is an
oscillating function of $z$, the distance of the atom from the
boundary. In order to find the total rate, we need to add the
Minkowski vacuum contribution, which can be obtained by setting
acceleration, $a$, to zero in the corresponding result given in
Ref.~\cite{ZYL06},  and the above boundary-dependent correction
term. The result is
\begin{eqnarray}
{\biggl\langle {d
H_A(\tau)\/d\tau}\biggr\rangle}_{tot}^{xx}&=&-{e^2\/3\pi
}\sum_{\omega_b>\omega_d}\omega_{bd}^4{|\langle
b|r_x(0)|d\rangle|}^2\,\biggl(1-{3\/16}f_x(z,\omega_{bd})\,\biggr)\;.\label{totx}
\end{eqnarray}
Obviously, with merely a substitution of $r_y$ for $r_x$ and
$f_y(z,\omega_{bd})$ =$f_x(z,\omega_{bd})$ for
$f_x(z,\omega_{bd})$, the above result also applies for the $yy$
component contributions, that is,
\begin{eqnarray}
{\biggl\langle {d
H_A(\tau)\/d\tau}\biggr\rangle}_{tot}^{yy}&=&-{e^2\/3\pi
}\sum_{\omega_b>\omega_d}\omega_{bd}^4{|\langle
b|r_y(0)|d\rangle|}^2\,\biggl(1-{3\/16}f_y(z,\omega_{bd})\,\biggr)\;.\label{toty}
\end{eqnarray}
Similarly, one has for the $zz$ component case that
\begin{eqnarray}
{\biggl\langle {d
H_A(\tau)\/d\tau}\biggr\rangle}_{b,vf}^{zz}&=&{e^2\/32\pi}\biggl(\sum_{\omega_b>\omega_d}
\omega_{bd}^4{|\langle
b|r_z(0)|d\rangle|}^2\,f_z(z,\omega_{bd})-\sum_{\omega_b<\omega_d}\omega_{bd}^4{|\langle
b|r_z(0)|d\rangle|}^2\,f_z(z,\omega_{bd})\biggr)\label{vfz2}\nonumber\\
\end{eqnarray}
for the contribution of vacuum fluctuations to the rate of change
of the atomic excitation energy, and
\begin{eqnarray}
{\biggl\langle {d
H_A(\tau)\/d\tau}\biggr\rangle}_{b,rr}^{zz}&=&{e^2\/32\pi}\biggl(\sum_{\omega_b>\omega_d}
\omega_{bd}^4{|\langle
b|r_z(0)|d\rangle|}^2\,f_z(z,\omega_{bd})+\sum_{\omega_b<\omega_d}\omega_{bd}^4{|\langle
b|r_z(0)|d\rangle|}^2\,f_z(z,\omega_{bd})\biggr)\label{rrz2}\nonumber\\
\end{eqnarray}
for that of radiation reaction, where function
$f_z(z,\omega_{bd})$ is given by
\begin{equation}
f_z(z,\omega_{bd})={4\/z^2\omega_{bd}^2}\cos(2z\omega_{bd})-{2\/z^3\omega_{bd}^3}\sin(2
z\omega_{bd})\;.
\end{equation}
It then follows that
\begin{eqnarray}
{\biggl\langle{dH_A(\tau)\/d\tau}\biggr\rangle}_{b,
tot}^{zz}={e^2\/16\pi}
\sum_{\omega_b>\omega_d}\omega_{bd}^4{|\langle
b|r_z(0)|d\rangle|}^2\,f_z(z,\omega_{bd})\;,\label{totzR}
\end{eqnarray}
and \begin{eqnarray}
{\biggl\langle{dH_A(\tau)\/d\tau}\biggr\rangle}_{tot}^{zz}&=&-{e^2\/3\pi}
\sum_{\omega_b>\omega_d}\omega_{bd}^4{|\langle
b|r_z(0)|d\rangle|}^2\biggl(1-{3\/16}\,f_z(z,\omega_{bd})\biggr)\;.\nonumber\\\label{totz}
\end{eqnarray}

After having presented all the results of our calculations, a few
comments are now in order.  First, although the presence of the
conducting boundary modifies both the vacuum fluctuations and
radiation reaction (refer, for example, to Eq.~(\ref{vfx2}) and
Eq.~(\ref{rrx2}) ), the effects of both contributions to the
spontaneous excitation rate, however, cancel exactly for an atom
in the ground state ($\omega_b<\omega_d$) (refer to
Eqs.~(\ref{totxR}, \ref{toty}, \ref{totzR}).  Therefore, the
presence of a plane boundary conspires to  modify  the vacuum
fluctuations and radiation reaction in such a way that the
delicate balance  between the vacuum fluctuations and radiation
reaction found in Ref.~\cite{ZYL06} in absence of boundaries
remains  and this ensures the stability of ground-state inertial
atoms in vacuum with a conducting boundary.  Second, if the atom
is polarized in the parallel direction, then, as the atom is
placed closer and closer to the boundary $(z\rightarrow0)$, the
rate of change of the atomic energy vanishes since
$f_x(z,\omega_{bd})$ and $f_y(z,\omega_{bd})$ approach zero. This
can be understood as a result of the fact that the tangential
components of the electric field vanish on the conducting plane.
However, if the polarization of the atom is along the normal
direction, then $f(z,\omega_{bd})\approx2$ when $(z\rightarrow0)$,
and one has
\begin{eqnarray}
{\biggl\langle{dH_A(\tau)\/d\tau}\biggr\rangle}_{tot}^{zz}\approx-{2e^2\/3\pi}
\sum_{\omega_b>\omega_d}\omega_{bd}^4{|\langle
b|r_z(0)|d\rangle|}^2\;.
\end{eqnarray}
This is two times that of the unbounded case and it can be
attributed to the fact that the reflection at the boundary doubles
the normal component of the fluctuating electric field. Thirdly,
for an atom which polarized in an arbitrary direction, we need to
add all contributions together ( Eq.~(\ref{totx}, \ref{toty},
\ref{totz}) ) and the result is
\begin{eqnarray}
{\biggl\langle {d
H_A(\tau)\/d\tau}\biggr\rangle}_{tot}=-{e^2\/3\pi}
\sum_{\omega_b>\omega_d}\omega_{bd}^4{|\langle
b|\textbf{r}(0)|d\rangle|}^2 -{e^2\/3\pi}
\sum_{\omega_b>\omega_d}{3\/16}\;\omega_{bd}^4{|\langle
b|r_i(0)|d\rangle|}^2f_i(z,\omega_{bd})\;.
\end{eqnarray}
Clearly the second term involving functions $f_i(z,\omega_{bd})$
 give the modifications induced by the presence of the boundary to the
rate of change of the mean atomic energy and they are oscillating
functions of $z$ with a modulated amplitude. Since $f_x$ and $f_y$
are different from $f_z$, the polarization of the atom in the
direction parallel to the boundary and that in the normal
direction are weighted differently in terms of their contributions
to the spontaneous emission rate of the atom.  Fourthly, When  the
distance of the atom from the boundary approaches infinity
$(z\rightarrow\infty)$,  all the oscillating functions approach
zero, and we recover the results of the unbounded case.
 Fifthly, let us note that the oscillating behavior of the
spontaneous emission rate of the hydrogen atom as a function of
$z$ may manifest itself in the intensity of the emission spectrum
and therefore might be verified in experiment.
 Take a typical transition frequency of a hydrogen atom,
$\omega_{bd}\sim10^{15}s^{-1}$,  for example, the amplitude of the
oscillating functions will show appreciable deviations from 1
 when $z\sim{c\/\omega_{bd}}\sim10^{-5}cm$, which is orders of magnitude larger than radius of the hydrogen atom.
Finally, the readers should be warned that our results are based
upon a particular of model for the hydrogen atom in which the
multipolar coupling between the atom and the electromagnetic
fields is assumed.

\section{Uniformly accelerated atom}

\subsection{Basic results}

 We now turn to the case in which the atom is uniformly
 accelerated in a direction parallel to the conducting plane
 boundary. We assume that the atom is at a distance $z$ from the
 boundary and is being accelerated in the $x$-direction with a proper
 acceleration $a$. Specifically, the atom's trajectory is described by
\begin{eqnarray}
t(\tau)={1\/a}\sinh a\tau\ ,\ \ \ x(\tau)={1\/a}\cosh a\tau\ ,\ \
\ y(\tau)=y_0,\ \ \ z(\tau)=z\;.\label{tra}
\end{eqnarray}
Let us introduce a unit vector pointing along the direction of
acceleration, $k^\mu=(0,1,0,0)$,  then the electric field
two-point function for the trajectory (\ref{tra}) can be evaluated
from its general form (\ref{EER}) in the frame of the atom with a
substitution $u=\tau-\tau'$ as follows
\begin{eqnarray}
\langle0|E_i(x(\tau))E_j(x(\tau'))|0\rangle_b
&=&-{a^4\/16\pi^2}{1\/
[\,\sinh^2{a\/2}(u-i\varepsilon)-a^2z^2\,]^3}\nonumber\\
&& \times\bigg\{\big[\,\delta_{ij}-2n_in_j+2az(n_ik_j+k_in_j)
\,\big]\sinh^2{au\/2}\nonumber\\
&& \quad+
a^2z^2\big[\,\delta_{ij}\cosh^2{au\/2}+(\delta_{ij}-2k_ik_j)\sinh^2{au\/2}\,\big]\,\bigg\}\;.
\label{eiej}
\end{eqnarray}
From Eq.~(\ref{eiej}), we obtain the Hadamard functions of the
field
\begin{eqnarray}
C_{ij}^F(x(\tau),x(\tau'))&=&-{a^4\/32\pi^2}\biggl({1\/[\,\sinh^2{a\/2}(u-i\varepsilon)-a^2z^2]^3
}+{1\/[\,\sinh^2{a\/2}(u+i\varepsilon)-a^2z^2\,]^3}\biggr)\nonumber\\
&&\times \bigg\{\big[\,\delta_{ij}-2n_in_j+2az(n_ik_j+k_in_j)
\,\big]\sinh^2{au\/2}\nonumber\\
&& \quad \quad +
\,a^2z^2\big[\,\delta_{ij}\cosh^2{au\/2}+(\delta_{ij}-2k_ik_j)\sinh^2{au\/2}\,\big]\,\bigg\}\;,\nonumber\\
\label{Hard}
\end{eqnarray}
and the Pauli-Jordan or Schwinger functions
\begin{eqnarray}
\chi_{ij}^F(x(\tau),x(\tau'))&=&-{ia\/16\pi
z}{\delta''\biggl(\sinh{au\/2}-az\biggr)-\delta''
\biggl(\sinh{au\/2}+az\biggr)\/
\sinh^2(au)-\cosh(au)\sinh^2{au\/2}+a^2z^2\cosh
(au)}\nonumber\\
&&\times\bigg\{\big[\,\delta_{ij}-2n_in_j+2az(n_ik_j+k_in_j)
\,\big]\sinh^2{au\/2}\nonumber\\
&& \quad \quad + \,a^2z^2\big[\,\delta_{ij}\cosh^2{au\/2}+(\delta_{ij}-2k_ik_j)\sinh^2{au\/2}\,\big]\,\bigg\}\;.\nonumber\\
\label{Pauli}
\end{eqnarray}
From the above expressions one can see that the only non-zero
components of the statistical functions are $xx,yy,zz,$ and $xz$
components. For an accelerating arbitrarily polarized atom, we
need to perform calculations for all non-zero statistical
functions in order to obtain the boundary-dependent contributions
of vacuum fluctuations and radiation reaction to the rate of
change of the atomic energy.


Let us now take the $xx$ component for example to show how the
calculations are to be carried out.  It follows from
Eq.~(\ref{Hard}) and Eq.~(\ref{Pauli}) that
\begin{eqnarray}
C_{xx}^F(x(\tau),x(\tau'))=-{a^4\/32\pi^2}\biggl({\sinh^2{au\/2}+a^2z^2\/
[\sinh^2{a\/2}(u-i\varepsilon)-a^2z^2]^3}+{\sinh^2{au\/2}+a^2z^2\/
[\sinh^2{a\/2}(u+i\varepsilon)-a^2z^2]^3}\biggr)\nonumber\\
\end{eqnarray}
and
\begin{eqnarray}
\chi_{xx}^F(x(\tau),x(\tau'))&=&-{ia\/16\pi
z}{\sinh^2{au\/2}+a^2z^2\/\sinh^2(au)-\cosh(au)\sinh^2{au\/2}+a^2z^2\cosh
(au)}\nonumber\\&&\times\biggl[\delta''\biggl(\sinh{au\/2}-az\biggr)-\delta''
\biggl(\sinh{au\/2}+az\biggr)\biggr]\;.
\end{eqnarray}
 The contributions of
vacuum fluctuations (\ref{hvf}) and radiation reaction (\ref{hrr})
to the rate of change of the mean atomic energy associated with
the above statistical functions can be written as
\begin{eqnarray} &&{\biggl\langle {d
H_A(\tau)\/d\tau}\biggr\rangle}_{b,vf}^{xx}={e^2a^4\/32\pi^2}\sum_d{|\langle
b|r_x(0)|d\rangle|}^2\;\omega_{bd}\nonumber\\
&& \quad\quad\times \int_{-\infty}^\infty
du\biggl({\sinh^2{au\/2}+a^2z^2\/
[\,\sinh^2{a\/2}(u-i\varepsilon)-a^2z^2\,]^3}+{\sinh^2{au\/2}+a^2z^2\/
[\,\sinh^2{a\/2}(u+i\varepsilon)-a^2z^2\,]^3}\biggr)e^{i\omega_{bd}u}\label{vfxx1}
\end{eqnarray}
and
\begin{eqnarray}
{\biggl\langle {d
H_A(\tau)\/d\tau}\biggr\rangle}_{b,rr}^{xx}&=&{iae^2\/16\pi
z}\sum_d{|\langle b|r_x(0)|d\rangle|}^2\;\omega_{bd}\nonumber\\
&& \quad \times\int_{-\infty}^\infty
du{\sinh^2{au\/2}+a^2z^2\/\sinh^2(au)-\cosh(au)\sinh^2{au\/2}+a^2z^2\cosh
(au)}\nonumber\\
&&
\quad\times\biggl[\delta''\biggl(\sinh{au\/2}-az\biggr)-\delta''
\biggl(\sinh{au\/2}+az\biggr)\biggr]
e^{i\omega_{bd}u}\;.\nonumber\\\label{rrxx1}
\end{eqnarray}
Here we have, as usual,  extended the range of integration to
infinity for sufficiently long times $\tau-\tau_0$. The integral
in Eq.~(\ref{vfxx1}) can be evaluated via the residue theorem to
get
\begin{eqnarray}
{\biggl\langle {d
H_A(\tau)\/d\tau}\biggr\rangle}_{b,vf}^{xx}&=&{e^2\/32\pi}\biggl[\sum_{\omega_b>\omega_d}\omega_{bd}^4{|\langle
b|r_x(0)|d\rangle|}^2f_{xx}(\omega_{bd},z,a)\biggl(1+{2\/e^{2\pi\omega_{bd}\/a}-1}\biggr)
\nonumber\\&&-\sum_{\omega_b<\omega_d}\omega_{bd}^4{|\langle
b|r_x(0)|d\rangle|}^2f_{xx}(\omega_{bd},z,a)\biggl(1+{2\/e^{2\pi|\omega_{bd}|\/a}-1}\biggr)\biggr]\;,\label{vfxx2}
\end{eqnarray}
where
\begin{eqnarray}
f_{xx}(\omega_{bd},z,a)&=&{2(1+4a^2z^2)\/z^2\omega_{bd}^2(1+a^2z^2)^2}\cos\biggl({2\omega_{bd}
\sinh^{-1}(az)\/a}\biggr)\nonumber\\&&+{4z^2\omega_{bd}^2-1-2z^2a^2
(1+2z^2a^2-2z^2\omega_{bd}^2)\/z^3\omega_{bd}^3(1+a^2z^2)^{5/2}}
\sin\biggl({2\omega_{bd}\sinh^{-1}(az)\/a}\biggr)\;.\label{g1}
\end{eqnarray}
A comparison of Eq.~(\ref{vfxx2}) with that of the unbounded
Minkowski space \cite{ZYL06} shows that, the boundary-dependent
contribution is in fact a ``nonthermal" correction proportional to
the oscillating function $f_{xx}(\omega_{bd},z,a)$.  With the help
of the following equations
\begin{eqnarray}
&&\delta(\sinh{au\/2}-az)={2\/a\cdot
\cosh[{1\/2}({au\/2}+\sinh^{-1}(az)~)~]}\;\delta\biggl(u-{2\/a}\sinh^{-1}(az)~\biggr)
\nonumber\\&&\delta(\sinh{au\/2}+az)={2\/a\cdot
\cosh[{1\/2}({au\/2}-\sinh^{-1}(az)~)~]}\;\delta\biggl(u+{2\/a}\sinh^{-1}(az)~\biggr)\;,\label{help}
\end{eqnarray}
we can calculate the contribution of the radiation reaction to the
rate of change of the atomic energy to get
\begin{eqnarray}
{\biggl\langle {d
H_A(\tau)\/d\tau}\biggr\rangle}_{b,rr}^{xx}={e^2\/32\pi}\biggl(&&\sum_{\omega_b>\omega_d}\omega_{bd}^4{|\langle
b|r_x(0)|d\rangle|}^2f_{xx}(\omega_{bd},z,a)\nonumber\\
&& +\sum_{\omega_b<\omega_d}\omega_{bd}^4{|\langle
b|r_x(0)|d\rangle|}^2f_{xx}(\omega_{bd},z,a)\biggr)\;.\nonumber\\\label{rrxx2}
\end{eqnarray}
Adding up the two contributions, (\ref{vfxx2}) and (\ref{rrxx2}),
we can get the total correction induced by the presence of the
boundary
\begin{eqnarray}
{\biggl\langle {d
H_A(\tau)\/d\tau}\biggr\rangle}_{b,tot}^{xx}&=&{e^2\/16\pi}\biggl[\sum_{\omega_b>\omega_d}\omega_{bd}^4{|\langle
b|r_x(0)|d\rangle|}^2f_{xx}(\omega_{bd},z,a)\biggl(1+{1\/e^{2\pi\omega_{bd}\/a}-1}\biggr)
\nonumber\\&&-\sum_{\omega_b<\omega_d}\omega_{bd}^4{|\langle
b|r_x(0)|d\rangle|}^2f_{xx}(\omega_{bd},z,a){1\/e^{2\pi|\omega_{bd}|\/a}-1}\;\biggr]\;.\label{totxxr}
\end{eqnarray}
The total rate of change of the atomic energy in the presence of a
conducting plane boundary can be obtained by further adding up the
Minkowski vacuum contribution given in Ref.~\cite{ZYL06}
\begin{eqnarray}
{\biggl\langle {d
H_A(\tau)\/d\tau}\biggr\rangle}_{tot}^{xx}&=&-{e^2\/3\pi}\biggl[\sum_{\omega_b>\omega_d}\omega_{bd}^4{|\langle
b|r_x(0)|d\rangle|}^2\biggl( 1 +
{a^2\/\omega_{bd}^2}-{3\/16}f_{xx}(\omega_{bd},z,a)\biggr)
\biggl(1+{1\/e^{2\pi\omega_{bd}\/a}-1}\biggr)
\nonumber\\&&-\sum_{\omega_b<\omega_d}\omega_{bd}^4{|\langle
b|r_x(0)|d\rangle|}^2\biggl( 1 +
{a^2\/\omega_{bd}^2}-{3\/16}f_{xx}(\omega_{bd},z,a)
\biggr){1\/e^{2\pi|\omega_{bd}|\/a}-1}\;\biggr]\;.\nonumber\\\label{totxx}
\end{eqnarray}

Similarly,  with the help of residue theorem and Eq.~(\ref{help}),
one can calculate the contributions related to other non-zero
components of the statistical functions and the results can be
summarized as follows. For a uniformly accelerated arbitrarily
polarized atom near a conducting plane,  the total
boundary-dependent contribution of vacuum fluctuations to the rate
of change of the mean atomic energy is given by
\begin{eqnarray}
{\biggl\langle {d
H_A(\tau)\/d\tau}\biggr\rangle}_{b,vf}&=&{e^2\/32\pi}\biggl[\sum_{\omega_b>\omega_d}\omega_{bd}^4
{|\langle b|r_i(0)|d\rangle|} {|\langle
d|r_j(0)|b\rangle|}f_{ij}(\omega_{bd},z,a)\biggl(1+{2\/e^{2\pi\omega_{bd}\/a}-1}\biggr)
\nonumber\\&&-\sum_{\omega_b<\omega_d}\omega_{bd}^4 {|\langle
b|r_i(0)|d\rangle|} {|\langle
d|r_j(0)|b\rangle|}f_{ij}(\omega_{bd},z,a)
\biggl(1+{2\/e^{2\pi|\omega_{bd}|\/a}-1}\biggr)\biggr]\;,\nonumber\\\label{vfyy2}
\end{eqnarray}
while for that of the radiation reaction, the result is
\begin{eqnarray}
{\biggl\langle {d
H_A(\tau)\/d\tau}\biggr\rangle}_{b,rr}&=&-{e^2\/32\pi}\biggl(\sum_{\omega_b>\omega_d}
\omega_{bd}^4 {|\langle b|r_i(0)|d\rangle|} {|\langle
d|r_j(0)|b\rangle|}f_{ij}(\omega_{bd},z,a)\nonumber\\
&& \quad -\sum_{\omega_b<\omega_d}
 \omega_{bd}^4
{|\langle b|r_i(0)|d\rangle|} {|\langle
d|r_j(0)|b\rangle|}f_{ij}(\omega_{bd},z,a)
\biggr)\;,\nonumber\\\label{rryy2}
\end{eqnarray}
where summation over repeated indices, $i,j$, is implied and
$f_{xx}(\omega_{bd},z,a)$ is given by Eq.~(\ref{g1}) while other
non-zero functions by
\begin{eqnarray}
f_{yy}(\omega_{bd},z,a)&=&{2(1+2a^2z^2)\/z^2\omega_{bd}^2(1+a^2z^2)}\cos\biggl({2\omega_{bd}
\sinh^{-1}(az)\/a}\biggr)\nonumber\\&&+{4z^2\omega_{bd}^2-1+4a^2z^4\omega_{bd}^2\/z^3\omega_{bd}^3(1+a^2z^2)^{3/2}}
\sin\biggl({2\omega_{bd}\sinh^{-1}(az)\/a}\biggr)\;,\label{f2}
\end{eqnarray}
\begin{eqnarray}
f_{zz}(\omega_{bd},z,a)&=&{2(2+a^2z^2+2a^4z^4)\/z^2\omega_{bd}^2(1+a^2z^2)^2}\cos\biggl({2\omega_{bd}
\sinh^{-1}(az)\/a}\biggr)\nonumber\\&&+{-2+a^2z^2(-5+4z^2\omega_{bd}^2)+4a^4z^6\omega_{bd}^2
\/z^3\omega_{bd}^3(1+a^2z^2)^{5/2}}
\sin\biggl({2\omega_{bd}\sinh^{-1}(az)\/a}\biggr)\;,\label{f4}
\end{eqnarray}
and
\begin{eqnarray}
f_{xz}(\omega_{bd},z,a)&=&{2a(-1+2a^2z^2)\/z\omega_{bd}^2(1+a^2z^2)^2}
\cos\biggl({2\omega_{bd}\sinh^{-1}(az)\/a}\biggr)\nonumber\\&&+{4az^2\omega_{bd}^2
+a+4a^3z^2(1+z^2\omega_{bd}^2)\/z^2\omega_{bd}^3(1+a^2z^2)^{5/2}}\sin
\biggl({2\omega_{bd}\sinh^{-1}(az)\/a}\biggr)\;.
\end{eqnarray}
As $z$, the distance of the atom from the boundary, approaches
infinity, all these functions approach zero and the
boundary-dependent contributions vanish as expected.  We  can add
the two contributions together to get the total contributions,
vacuum fluctuations plus radiation reaction, to the rate of change
of the atomic energy induced by the presence of the conducting
plane
\begin{eqnarray}
{\biggl\langle {d H_A(\tau)\/d\tau}\biggr\rangle}_{b,tot}
&=&{e^2\/32\pi}\biggl[\sum_{\omega_b>\omega_d}\omega_{bd}^4
{|\langle b|r_i(0)|d\rangle|} {|\langle
d|r_j(0)|b\rangle|}f_{ij}(\omega_{bd},z,a)\biggl(1+{1\/e^{2\pi\omega_{bd}\/a}-1}\biggr)
\nonumber\\&&-\sum_{\omega_b<\omega_d}\omega_{bd}^4 {|\langle
b|r_i(0)|d\rangle|} {|\langle
d|r_j(0)|b\rangle|}f_{ij}(\omega_{bd},z,a)
{1\/e^{2\pi|\omega_{bd}|\/a}-1}\,\bigg]\;.\nonumber\\
\end{eqnarray}
It follows immediately that the total rate of change of the atomic
energy with the Minkowski vacuum term \cite{ZYL06} included is
\begin{eqnarray}
&&{\biggl\langle {d
H_A(\tau)\/d\tau}\biggr\rangle}_{tot}\nonumber\\&&\quad=-{e^2\/3\pi}\biggl[\sum_{\omega_b>\omega_d}\omega_{bd}^4
{|\langle b|r_i(0)|d\rangle|} {|\langle
d|r_j(0)|b\rangle|}\biggl(\big(1 +
{a^2\/\omega_{bd}^2}\big)\delta_{ij}-{3\/16}f_{ij}(\omega_{bd},z,a)\biggr)
\biggl(1+{1\/e^{2\pi\omega_{bd}\/a}-1}\biggr)
\nonumber\\
&&\quad-\sum_{\omega_b<\omega_d}\omega_{bd}^4{|\langle
b|r_i(0)|d\rangle|} {|\langle d|r_j(0)|b\rangle|} \biggl(\big(1 +
{a^2\/\omega_{bd}^2}\big)\delta_{ij}-{3\/16}f_{ij}(\omega_{bd},z,a)\biggr){1\/e^{2\pi|\omega_{bd}|\/a}-1}\;\biggr]\;.\nonumber\\\label{Finaltot}
\end{eqnarray}

\subsection{Comments and discussions }
 A few comments and discussions are now in order for results obtained in the proceeding subsection.
It is interesting to note that Eq.~(\ref{Finaltot}) reveals that
for an accelerated atom in the ground state $(\omega_b <
\omega_d)$, the effects of both contributions do not exactly
cancel as in the case of an inertial atom, so the delicate balance
between the vacuum fluctuations and radiation reaction no longer
exists if the atom is accelerated, although both contributions of
the vacuum fluctuations and radiation are altered for accelerated
atoms in the presence of the boundary as contrasted with the case
without boundaries. There is a positive contribution from the
$\omega_{b}< \omega_{d}$ term, therefore transitions of
ground-state accelerated atoms to excited states are allowed to
occur in vacuum with boundaries. The presence of the boundary
modulates the transition rate with the functions,
$f_{ij}(\omega_{bd},a,z)$ and makes the rate a function of $z$,
the atom distance from the boundary. It is interesting to note
that the boundary-induced contribution is effectively a nonthermal
correction, thus depending the atom's distance from the boundary,
the nonthermal correction (the term proportional to $a^2$) which
is already present in the unbounded case may get enhanced or
weakened by the presence of the boundary. This nonthermal effect
which appears even when the boundary is absent may become
appreciable for observation when the acceleration is of order
necessary to observe the Unruh effect in atomic systems
\cite{Rosu}. With the presence of the boundary, the nonthermal
effect is expected to be enhanced for atoms on certain
trajectories and thus more likely to be observed. For a given atom
with a certain polarization, a typical transition frequency and a
certain acceleration $a$, one can find a value of $z$  where the
nonthermal correction induced by the presence of the boundary is
comparable with that already present without boundaries. For
example, if the atom is polarized in the $z$-direction, then for a
typical transition frequency of a hydrogen atom,
$\omega_{bd}\sim10^{15}s^{-1}$,  and an acceleration,
$a\sim10^{25}cm/s^2$, typical acceleration for the Unruh effect to
be observable in atomic systems, one can show that this value of
$z$ is $z\sim10^{-5}cm$.

 At the same time, it is interesting to note that,
, for an accelerated atom which is only polarized in the $x$ or
$y$ or $z$ direction, there exists a certain value of  $z$ for
every pair of $a$ and $\omega_{bd}$, such that
${a^2\/\omega_{bd}^2}-{3\/16}f_{ii}(\omega_{bd},z,a)=0$,  that is,
for atoms accelerated on the trajectory characterized by this
value of $z$,  the nonthermal corrections vanish.

Let us now note that as the acceleration, $a$, approaches zero,
one has
\begin{eqnarray}
f_{xx}(\omega_{bd},z,a)&=& f_x(z,\omega_{bd})+\biggl(
{13-4z^2\omega_{bd}^2\/3\omega_{bd}^2}\cos(2z\omega_{bd})+{3-32z^2\omega_{bd}^2\/6z\omega_{bd}^3}\sin(2z\omega_{bd})\bigg)a^2+O[a]^4\;,
\nonumber\\ \label{fxx}
\end{eqnarray}
\begin{eqnarray}
f_{yy}(\omega_{bd},z,a)&=& f_y(z,\omega_{bd})+\biggl(
{7-4z^2\omega_{bd}^2\/3\omega_{bd}^2}\cos(2z\omega_{bd})+{9-32z^2\omega_{bd}^2\/6z\omega_{bd}^3}\sin(2z\omega_{bd})\bigg)a^2+O[a]^4\;,
\nonumber\\\;,
\end{eqnarray}
\begin{eqnarray}
f_{zz}(\omega_{bd},z,a)&=&
f_z(z,\omega_{bd})+\biggl({16z\/3\omega_{bd}}\sin(2z\omega_{bd})-{16\/3\omega_{bd}^2}\cos(2z\omega_{bd})\bigg)a^2
+O[a]^4\;,
\end{eqnarray}
and
\begin{eqnarray}
f_{xz}(\omega_{bd},z,a)=\biggl({4z^2\omega_{bd}^2-1\/z^2\omega_{bd}^3}\sin(2z\omega_{bd})-{2\/z\omega_{bd}^2}\cos(2z\omega_{bd})\bigg)a
+O[a]^3\;.
\end{eqnarray}
This shows that the rate of change of the mean atomic energy will
be that for an inertial atom found in the preceding section plus
an acceleration-dependent correction, and if the acceleration
equals zero, we recover the result of Section III.

We now examine what happens as the atom is placed closer and
closer to the boundary ($z\rightarrow 0$). In this case, one finds
for any finite acceleration,
 $a$,  that
\begin{eqnarray}
1+{a^2\/\omega_{bd}^2}-{3\/16}f_{xx}(\omega_{bd},z,a)=
\bigg(4a^2+{16a^4\/5\omega_{bd}^2}+{4\omega_{bd}^2\/5}\bigg)z^2 +
O[z]^4\;,\label{fxxz0}
\end{eqnarray}
\begin{eqnarray}
1+{a^2\/\omega_{bd}^2}-f_{yy}(\omega_{bd},z,a)=
\bigg(2a^2+{6a^4\/5\omega_{bd}^2}+{4\omega_{bd}^2\/5}\bigg)z^2 +
O[z]^4\;,\label{fyyz0}
\end{eqnarray}
\begin{eqnarray}
1+{a^2\/\omega_{bd}^2}-f_{zz}(\omega_{bd},z,a)=2\bigg(1+{a^2\/\omega_{bd}^2}\bigg)
-\bigg(2a^2+{18a^4\/5\omega_{bd}^2}+{2\omega_{bd}^2\/5}\bigg)z^2 +
O[z]^4\;,
\end{eqnarray}
and
\begin{eqnarray}
f_{xz}(\omega_{bd},z,a)=\biggl({32a\/3}+{32a^3\/\omega_{bd}^3}\bigg)z
+O[z]^3\;.
\end{eqnarray}
Therefore, if the atom is polarized in a direction parallel to the
conducting plane, then the spontaneous excitation rate of the atom
diminishes to zero quadratically in $z$ (refer to
Eq.~(\ref{fxxz0}) and Eq.~(\ref{fyyz0}) ) as the boundary is
approached ($z\rightarrow 0$). Recall the result in the proceeding
Section, we see that the fact that the total excitation rate
vanishes on the boundary is independent of whether the atom is
accelerated or in uniform motion and this can be understood as a
result of the fact that the tangential components of the electric
field vanish on the conducting plane. However, two parallel
directions, the $x$-direction ( along the acceleration) and the
$y$-direction (perpendicular to the acceleration), are now not
equivalent as in the inertial case, since $f_{xx}$ and $f_{yy}$
are not equal. On the other hand, if the atom's polarization is
perpendicular to the conducting plane, then as $z$, distance of
the atom from the boundary, approaches zero, we obtain
\begin{eqnarray}
{\biggl\langle{dH_A(\tau)\/d\tau}\biggr\rangle}_{tot}^{zz}&=&-{2e^2\/3\pi}
\biggl[\sum_{\omega_b>\omega_d}\omega_{bd}^4{|\langle
b|r_z(0)|d\rangle|}^2\biggl({a^2\/\omega_{bd}^2}+1\biggr)
\biggl(1+{1\/e^{2\pi\omega_{bd}\/a}-1}\biggr)
\nonumber\\&&-\sum_{\omega_b<\omega_d}\omega_{bd}^4{|\langle
b|r_z(0)|d\rangle|}^2\biggl({a^2\/\omega_{bd}^2}+1
\biggr){1\/e^{2\pi|\omega_{bd}|\/a}-1}\;\biggr]\;,
\end{eqnarray}
which is just two times the corresponding result in an unbounded
Minkowski space \cite{ZYL06}. This enhancement can be attributed
to the fact that the reflection at the boundary doubles the normal
component of the fluctuating electric field. The above analysis
tells us that even if the atom is isotropically polarized, each of
three equal polarizations will be weighted differently in terms of
its contribution to the rate of change of the mean atomic energy.

Finally,  another interesting feature to be noted is that if the
 polarization of the atom is in $x-z$ plane, the rate of change of
 the atomic energy gets an extra contribution associated with
 $f_{xz}$ as compared with the inertial case. This extra
 contribution vanishes when $a$ goes to zero and we recover the
 result of the inertial case as expected. Also, as $z$, the distance of the atom to the
boundary, approaches zero or infinity, the  contribution
diminishes to zero too.

\section{Conclusions}

In conclusion, assuming a multipolar coupling between a
multi-level atom and a quantum electromagnetic field,  we have
studied the spontaneous emission and absorption of  both an
inertial and a uniformly accelerated atom near a conducting plane
in vacuum and separately calculated the contributions of vacuum
fluctuations and radiation reaction to the rate of change of the
atomic energy.

In the case of an inertial atom,  our results show that both the
contributions of vacuum fluctuations and radiation reaction to the
rate of change of the atomic energy are  modified by the presence
of the boundary, but the balance between them remains for ground
state atoms and  this ensures the atom's
 stability in its ground state. The spontaneous emission rate of the atom in this
 case is an oscillating function of atom's distance from the
 boundary and this oscillating behavior may offer a possibility
for experimental test.

 If the atom moves with constant proper acceleration, the perfect balance between
 the contributions of vacuum fluctuations and radiation reaction
that ensures the stability of ground-state atoms is disturbed,
making spontaneous transition of ground-state atoms to excited
states possible in vacuum with a conducting boundary.  The
presence of the boundary modulates the spontaneous absorption rate
with functions dependent on the acceleration and the atom's
distance from the boundary. The boundary-induced contribution is
effectively a nonthermal correction, which enhances or weakens the
nonthermal effect already present in the unbounded case, thus
possible making the effect easier to observe. The appearance of
nonthermal correction terms suggest that, the effect of
electromagnetic vacuum fluctuations is not totally equivalent to
that of a thermal field as is the case for a scalar field in the
unbounded space~\cite{Audretsch94}. However, it is interesting to
note that, for atoms on some particular trajectories, the
nonthermal correction induced by the presence of the boundary and
that already present in the unbounded case may cancel. The
calculations performed in this paper also tell us that each of
three polarizations of the atom is weighted differently in terms
of its contribution to the rate of change of the mean atomic
energy even if the atom is isotropically polarized, as a result of
the anisotropy of the configuration due to the presence of the
boundary and the atom's acceleration.


\begin{acknowledgments}
This work was supported in part  by the National Natural Science
Foundation of China  under Grants No.10375023 and No.10575035, the
Program for NCET (No.04-0784), the Key Project of Chinese Ministry
of Education (No.205110) and the Research fund of Hunan Provincial
Education Department (No. 04A030)
\end{acknowledgments}


\end{document}